%Paper: astro-ph/9508084
%From: pavel@auriga.ari.uni-heidelberg.de (Pavel Kroupa)
%Date: Fri, 18 Aug 1995 15:20:35 +0100

\raggedbottom

% Definitions for symbols:
\def\e-p{$e-{\rm log}_{10}P$}
\def\P{ {\rm log}_{10}P }
\def\nex{\par\noindent\hang}

\null
\vskip 20mm
\centerline{\bf The Dynamical Properties of Stellar Systems in the Galactic
Disc}

\vskip 2mm

\centerline{\bf K2}

\vskip 10mm
\centerline{\bf Pavel Kroupa}
\vskip 10mm
\centerline{Astronomisches Rechen-Institut}
\vskip 2mm
\centerline{M{\"o}nchhofstra{\ss}e~12-14, D-69120~Heidelberg, Germany}
\vskip 10mm
\centerline{e-mail: S48@ix.urz.uni-heidelberg.de}

\vfill

\centerline{MNRAS, in press}

\vfill\eject

\hang{
\bf Abstract.
\rm
We postulate that stars in the Galactic field are born in aggregates of binary
stars with half mass radii $R_{0.5}$ and number of binaries $N_{\rm bin}$ which
are dynamically equivalent to the dominant mode cluster
$(N_{\rm
bin},R_{0.5})\approx(200,0.8\,{\rm pc})$.
Binary orbits are distributed according to an initial
period distribution which is consistent with pre-main sequence data.
Stellar masses are paired at random from the KTG(1.3) mass function.
We develop a simple
model for the redistribution of orbital angular momentum and energy
in short-period proto-binary systems ({\it pre-main sequence
eigenevolution}), which establishes the observed correlations between
eccentricity, mass ratio and period. The
evolution of orbital parameters owing to perturbations by neighbouring systems
({\it stimulated evolution}) within the dominant mode cluster
places 1-2~per cent of all orbits into the {\it eigenevolution region}
($P<100\,{\rm days}, e>0.1$ approximately) of
the eccentricity-period diagram. The number of such {\it forbidden} orbits at
any
time is a function of the stellar number density, the dynamical and the nuclear
age of the cluster.
Observations of binaries in
clusters should reveal the odd binary with {\it forbidden orbital parameters}.
Examples
of such systems may be the pre-main sequence binaries P2486 and EK~Cep and
binaries in stellar clusters with eccentric orbits at periods
smaller than the circularisation cutoff period.
Eigenevolution is expected to depopulate the eigenevolution region
within $10^5$~yrs for pre-main sequence binaries, but main
sequence binaries with forbidden orbits should remain in the eigenevolution
region for times of order $10^9$~yrs.
We show that the binary star population must have
a birth eccentricity distribution which is approximately
dynamically
relaxed because stimulated evolution in the dominant mode cluster cannot
sufficiently thermalise a significantly different distribution. After
disintegration of the dominant mode cluster we have
a population of Galactic field systems with orbital parameters as observed,
with a surviving proportion of binaries of $f_{\rm tot}=0.48\pm0.03$ which
compares favourably with the observed proportion $f_{\rm tot}^{\rm
obs}=0.47\pm0.05$. The rise of the period
distribution to a maximum at log$_{10}P\approx4.8$ reflects approximately
the initial distribution, whereas the decay for log$_{10}P>4.8$ results from
stimulated evolution. The mass-ratio distribution of G~dwarf binaries is
depleted at small mass ratios and has the shape of the main sequence
distribution despite initially being the KTG(1.3)
mass function. We predict and tabulate the mass ratio
distribution for main sequence binaries with a primary star less massive than
$1.1\,M_\odot$. Our model Galactic field
stellar population has a binary proportion among G-, K- and M-dwarfs in good
agreement with the observational data.
Too few triple and quadruple systems form by capture to
account for the number of observed systems. An example of a triple
system which may have formed by capture in the birth aggregate may be
Proxima~Cen--$\alpha$~Cen~A/B.
We compare the specific angular momentum distribution of our initial binary
star population with the observed distribution of specific angular momenta of
molecular cloud cores. According to our model about 40~per cent of all
late-type stars are single after cluster dissolution but had companions with
log$_{10}P\ge6$ at birth. These stars are expected to have circumstellar disks
(and possible planetary systems) extending to at least about 40~AU.

\vskip 5mm

}

\vfill\eject

\noindent{\bf 1 INTRODUCTION}
\vskip 12pt
\noindent
The {\it dyanamical properties of stellar systems} are the stellar mass
function,
the proportion of binaries (and more generally of multiple systems) and their
orbital parameters. In depth analysis of star count data provides the Galactic
field stellar mass function (Kroupa, Tout \& Gilmore 1993). The distribution of
the remaining dynamical properties of stellar systems in the Galactic disk can
be derived under simple assumptions about the initial distribution of the
dynamical properties if the majority of low-mass stars form in
aggregates which are
{\it dynamically equivalent} to the {\it dominant mode cluster} (Kroupa 1995a,
hereafter
paper K1). The simple assumptions are consistent with data on pre-main sequence
stars and are: (i) all stars are born in binary systems; (ii) which have
uncorrelated component masses. Quantifying {\it stimulated evolution} (i.e. the
evolution of orbital parameters owing to near-neighbour perturbations) allows
K1 to derive an initial distribution of periods of the binary stars
(equation~11 in K1):

$$ f_{\rm P,in} = \eta \, { \left({\rm log}_{10}P -
{\rm log}_{10}P_{\rm min}\right)  \over \delta + \left({\rm log}_{10}P -
{\rm log}_{10}P_{\rm min}\right)^2},  \eqno (1)$$

\noindent where $P\ge P_{\rm min}$ is the orbital period in days and $\eta$ and
$\delta$ are parameters which determine the shape and the maximum period which
must satisfy log$_{10}P_{\rm max}>7.5$ in order to account for observed systems
with such large periods. K1 adopts $\eta=3.50, \delta=100$ and $P_{\rm
min}=1\,$day so that log$_{10}P_{\rm max}=8.78$ because by our initial
assumption that all stars form in binary systems equation~1 must have an
integral over period equal to unity (equations~9-11 in K1). The initial
distribution of eccentricities
is taken to be thermal but is not critical for the conclusions of K1. The
dominant mode cluster is assumed to be in virial equilibrium initially and is
approximated by $(N_{\rm
bin},R_{0.5})\approx(200,0.8\,{\rm pc})$, where $N_{\rm bin}$ is the number of
binaries distributed according to a Plummer density law with half-mass radius
$R_{0.5}$. This is consistent with direct imaging surveys of star-forming
regions which identify similar structures to probably be the dominant mode of
star formation (see K1 and references therein).

This paper (herafter K2) is the second in a series of three papers K1, K2 and
K3. In K1 we show that inverse dynamical population synthesis
implies that clustered star formation may be the dominant mode of star
formation. In K3 (Kroupa 1995b)
we study in detail the dynamical evolution of the stellar
aggregates simulated in K1 and K2. In this paper (K2) we study in detail the
dynamical properties of stellar systems that result if stars are born
in the dominant mode cluster.

In order to study the effects of stimulated evolution on the short period
(log$_{10}P<3$) binary star population we need to develop a model of the
eccentricity--period diagram which reproduces its observational features. These
include circular orbits for $P<P_{\rm c}$, where the circularisation cutoff
period is $P_{\rm c}\approx12$~days for main sequence stars in the Galactic
field, and the
absence of eccentric orbits at small periods. The basic assumption of our model
is that star formation is not responsible for the
difference between the eccentricity and mass ratio distributions at
$\P<3$ and $\P>3$ (see figure~2 in K1), but that this difference is due to
the evolution of the orbital parameters in short period systems because of the
interaction between the two accreting proto stars. We refer to this evolution
as {\it pre-main sequence eigenevolution}. By modelling the initial \e-p
diagram we can hope to understand better the nature of `anomalous' pre-main
sequence and main sequence orbits which may appear in this diagram.
We need to modify $\eta$ and $\delta$ in equation~1 to account for the
redistribution of angular momentum and energy in short period systems owing to
pre-main sequence eigenevolution.

We assume stars form in the dominant mode cluster, $(N_{\rm
bin},R_{0.5})=(200,0.85\,{\rm pc})$, listed under
p.K2,K3 in table~1 of K1 and perform $N_{\rm run}=20$ simulations to improve
statistical significance of our results. This cluster has an initial central
number density of 320~stars~pc$^{-3}$, a mass of $128\,M_\odot$, a crossing
time $T_{\rm cr}=3.5$~Myrs and a median relaxation time $T_{\rm relax}=11$~Myrs
(table~1 in K1). For the numerical integration of stellar orbits we use the
N-body program {\it nbody5} Aarseth (1994) has developed. A standard Galactic
tidal field is modelled. The average stellar number density in the central 2~pc
sphere decays to less than 0.1~stars~pc$^{-3}$ after $740\pm150$~Myrs when the
cluster has completely disintegrated. We retain all stars, irrespective of
whether they are bound to the intial aggregate or not. All final distributions
are evaluated
at time $t=1$~Gyr, i.e. well after cluster disintegration. Our method of
finding all bound binary systems at any given time is described in K1. We use
the same definition for the distribution of periods, $f_{\rm P,i}({\rm
log}_{10}P,t)$, as K1 (equation~7 in K1). It is the number of
orbits in the ith log$_{10}P$ bin divided by the total number of stellar
systems. The
proportion of binary systems, $f_{\rm tot}$, is the sum of $f_{\rm P,i}$ over
all periods (equations~2 and~7 in K1). We follow K1 and only consider stars
less massive than $1.1\,M_\odot$ and use the initial stellar mass
function derived by Kroupa et al. (1993). It is conveniently approximated by
$\xi(m)\propto m^{-\alpha_i}$, where $0.70\le\alpha_1\le1.85$ for
$0.1\,M_\odot\le m<0.5\,M_\odot$, $\alpha_2=2.2$ for $0.5\,M_\odot\le
m<1\,M_\odot$, $\alpha_3=2.7$ for $1\,M_\odot\le m$, and $\xi(m)\,dm$ is the
number of stars in the mass range $m$ to $m+dm$. From here on we refer to
$\xi(m)$ as the KTG($\alpha_1$) mass function and use $\alpha_1=1.3$ (K1).
We assume
component masses in binary systems are not correlated at birth.

We emphasise that if
the suggestion by K1 is true that most stars may form in stellar aggregates
that
are dynamically equivalent to our dominant mode cluster then these aggregates
must dissolve on a timescale less than 10~Myrs (see K1) which is
significantly
shorter than the evolution timescale of our dominant mode cluster. The presumed
dominant mode embedded cluster probably expels most binding mass within
10~Myrs. After dissolution of the dominant mode embedded cluster we expect to
obtain a very similar distribution of dynamical properties of stellar systems
as we obtain here (dynamical equivalence: K1, K3).

The assumptions about star formation made here are not complete in so far as
the birth of massive stars, nor of higher order multiple systems, nor a
changing background
potential in the young stellar aggregate are modelled. These will be future
extensions of the present work.

In Section~2 we introduce our model for eigenevolution and in Section~3 we
compare the dynamical properties of our model Galactic field stellar population
with the observational constraints. The initial orbital
parameters of our model binary star population are compared with the specific
angular momentum distribution of molecular cloud cores in Section~4, and
Section~5 contains a discussion of circumstellar disks. Finally, Section 6
concludes this paper.

\bigskip
\bigbreak
\noindent{\bf 2 EIGENEVOLUTION}
\nobreak\vskip 10pt\nobreak
\noindent In this section we device a model for the correlation between
eccentricity and mass ratio with period. According to this model angular
momentum and energy are redistributed in only about 10--20~per cent of all
pre-main
sequence binary systems, namely those with $P<10^3$~days and those with longer
periods and eccentric orbits at birth.

\nobreak\vskip 10pt\nobreak
\noindent{\bf 2.1 Introductory remarks}
\nobreak\vskip 10pt\nobreak
\noindent Stimulated evolution cannot account
for the
lack of orbits in the large eccentricity--small period region seen in all
observational eccentricity--period diagrams of pre-main sequence binaries.
An \e-p diagram of our orbits at any time is a
rectangular region given by $0\le e\le1$ and ${\rm log}_{10}P_0\le {\rm
log}_{10}P\le {\rm log}_{10}P_1$, where $P_0\approx1\,$days and
$P_2\approx10^9\,$days are not equal to $P_{\rm min}$ and $P_{\rm max}$
(equation~1),
respectively, because stimulated evolution forces a few orbits outside this
range.  However,
the distribution of observational data in any currently available $e-{\rm
log}_{10}P$ diagram shows
that for log$_{10}P<3$ there is an absence of systems with large
eccentricity which is established at a very early age (at the latest by about
1~Myr, Mathieu 1994). The thick long-dashed line in the topmost panel in
Fig.~1 approximately represents this upper envelope of the main-sequence
data (Duquennoy
\& Mayor 1991), and the birth distribution (in Section~2.5 we discuss the
difference between {\it birth} and {\it initial} distributions) of our model
systems clearly shows
a discrepancy with the observational data.
The model distribution of eccentricities is in
acceptable agreement
with the observations for log$_{10}P>3$, but in bad agreement for log$_{10}P<3$
(central panel in Fig.~1). The birth distribution of periods we adopt shows
there are too few model systems with $P<100\,$days (bottom panel in Fig.~1).

Pre-main sequence stars have larger radii than main sequence stars of the same
mass, and significant interaction between the
two stars can be expected if the periastron distance is comparable to their
radii. For example, a $1\,M_\odot$ star at the birthline has a radius of about
$5\,R_\odot$ (see Zahn \& Bouchet 1989 and references therein). In a binary
consisting of two such stars with a period of
10~days
eccentricities larger than about~0.6 would imply collision of both components
at periastron. Orbits with smaller eccentricities must evolve at a rate that
depends on the ratio of the pre-main sequence stellar radius and the periastron
distance. Orbital energy and orbital angular momentum are redistributed within
this system at a rate which depends on the tidal deformation of the stars and
on their internal structure, composition and rotation. The rate of tidal
dissipation of kinetic energy
is different for stars with radiative and convective envelopes, being much
enhanced in the latter (Zahn
1992). In the present context the application by
Zahn \& Bouchet (1989) of
tidal circularisation theory to post-birthline pre-main sequence
binaries descending along the Hayashi track is interesting. They suggest that
tidal
circularisation completes by 0.1~Myrs, i.e. on a very short time scale, because
of the strong dependence of tidal dissipation on the stellar radius and the
deep convection zones of pre-main sequence stars on the Hayashi track. We refer
the reader to the more thorough discussion
of tidal circularisation by Mathieu et al. (1992).

We do not adopt any of the
proposed theories to describe the early evolution of short period orbits
because
the theories do not easily account for evolution if the initial eccentricity is
larger than about 0.3. Also, we expect that significant evolution of the
orbital elements occurs before the proto-stars reach the birth line.
Major complications owing to continued accretion onto the proto-stars and
dynamical friction in the circum-protostellar material make pre-main sequence
``tidal circularisation'' inaccessible to rigorous theoretical investigation.
Direct
hydrodynamical simulations cannot help because they cannot presently be
extended beyond a few orbital periods at which stage most of the
circum-protostellar material remains to be accreted. The resulting
orbital parameters after the stellar masses have grown to approximately the
final size thus remain inaccessible (Bonnell 1994).

Given this state of affairs we shall from here on refer to the evolution of
orbital elements owing to system-internal redistribution of orbital energy and
angular momentum in the proto-stellar binary as {\it pre-main sequence
eigenevolution} rather than tidal
circularisation.

We cannot model eigenevolution
self consistently because this task (i.e. the interaction of the proto-stars
and the subsequent changes to the phase-space variables) lies beyond the scope
of the present investigation. We also do not model the possible increase in
eccentricity owing to resonances with gaseous disks because eccentricity
driving (Lubow \& Artymowicz 1992) is difficult to verify observationally
(Mathieu 1994). Instead, we assume proto-binary systems are born with periods
distributed according to equation~8 below, with eccentricities distributed
according to the thermal distribution (equation~4 in K1), and uncorrelated
component masses. Prior to starting integration with nbody5 we evolve this
distribution of birth orbital elements to an initial distribution according to
our model of eigenevolution below, which was inspired by Zahn
(1977) and Duquennoy, Mayor \& Mermilliod (1992). This process models the very
short (probably $<10^5$~yrs) duration of pre-main sequence eigenevolution
during the proto-stellar phase
so that the orbital parameters of short period binary systems are established
during the proto-stellar accretion phase.
According to our model, pre-main sequence
eigenevolution empties the \e-p diagram of eccentric orbits with short period.
We refer to the region above the thick long dashed curve
in the top panel of Fig.~1 as the {\it eigenevolution region} of the diagram.
However, while our model correctly reproduces the observed correlations of
orbital parameters, we caution that our eigenevolution model is rather crude,
i.e. it is primarily phenomenological in nature.

This approach allows us to obtain an insight into the effects stimulated
evolution has on the distribution of orbits in this diagram. Any orbits found
in the eigenevolution region subsequent to starting the numerical integration
we refer to as having {\it forbidden orbital parameters} or simply as {\it
forbidden orbits}. If stimulated evolution places a pre-main sequence binary
with components on the Hayashi track into the
eigenevolution region then we expect eigenevolution to rapidly (within about
$10^5$~yrs) circularise the forbidden orbit (Zahn \& Bouchet 1989).
However, we are interested to learn how many forbidden orbits have
been created, and so we do not eigenevolve these until we stop the N-body
integration. Thus our simulation is not self-consistent because we do not take
account of tidal dissipation on the equations of motion of each star and have
to leave a binary with a forbidden orbit where stimulated evolution placed it
in the \e-p diagram. Subsequent stimulated evolution of such an orbit is very
unlikely though, because these hard binaries have small interaction cross
sections, and so our approximation (i.e. pre-main sequence
eigenevolution then stimulated evolution and finally main sequence
eigenevolution) is reasonable for our purpose. Aarseth (1994 and private
communication) is incorporating tidal dissipation in his N-body code so that a
consistent treatment of a forbidden orbit will become possible in the future.

\nobreak\vskip 10pt\nobreak
\noindent{\bf 2.2 The model}
\nobreak\vskip 10pt\nobreak
\noindent
We assume significant eigenevolution occurs when the accreting proto-stars are
at periastron

$$R_{\rm peri} =  (1-e)\,P_{\rm yrs}^{2\over3}\,(m_1+m_2)^{1\over3},
\eqno(2)$$

\noindent where $P_{\rm yrs}=P/365.25$ is the period in years.
We assume that if the binary system is born with the birth
eccentricity
$e_{\rm birth}$ then the system evolves according to

$$ {1\over e}{de\over dt} = -\rho'    \eqno(3a)$$

\noindent
to the final eccentricity given by

$${\rm log}_{\rm e}e_{\rm fin} = -\rho + {\rm log}_{\rm e}e_{\rm birth},
\eqno(3b)$$

\noindent where

$$\rho = \int^{\Delta t}_0 \rho'\,dt= \left({\lambda\,R_\odot \over
R_{\rm peri}}\right)^\chi,
\eqno(4)$$

\noindent and we assume $R_{\rm peri}$ is constant (compare with Goldman \&
Mazeh 1994);
$R_\odot=4.6523\times10^{-3}\,$AU is the
radius of the sun; $\Delta t$ is the time interval after which
pre-main sequence eigenevolution
becomes insignificant ($\Delta t<10^5$~yrs approximately, Zahn \& Bouchet 1989)
and $\rho'^{-1}$ is the circularisation or pre-main sequence
eigenevolution timescale (compare with Duquennoy, Mayor
\& Mermilliod 1992). The period evolves to

$$P_{\rm fin} = P_{\rm birth}\,\left({m_{\rm tot,birth}\over m_{\rm
tot,fin}}\right)^{1\over2}\,\left({1-e_{\rm birth} \over 1-e_{\rm
fin}}\right)^{3\over2}, \eqno(5)$$

\noindent where $m_{\rm tot,birth}$ and $m_{\rm tot,fin}$ are, respectively,
the birth and final sum of the masses of the two companions.
We merge two close proto stars
if their semimajor axis after pre-main sequence eigenevolution
is $a\le10\,R_\odot$. This is a very simple criterion for merging, and a
different scenario based on a criterion that depends on eccentricity as well,
such as merging if the periastron distance is smaller than a few stellar raii,
might be more realistic. However, as it is not
our present aim to model merging during the proto-stellar accretion phase,
but merely to acknowledge the presence of merged binaries at the birth line and
to ease the computational burden of integrating the equations of motion, we
have chosen the
simple criterion above, and have a dopted the somewhat arbitrary factor of~10.

Choosing appropriate dimensionless parameters $\lambda$ and $\chi$ allows us
to model the distribution of systems in the $e-{\rm log}_{10}P$
diagram. Basically, $\lambda$ measures the length scale over which significant
evolution of the orbital elements during the proto stellar phase occurs, and
$\chi$ measures the `interaction strength' between the two proto stars in the
binary system.

{}From the discussion in section~2 of K1
we know that the mass ratio distribution for short period solar-type binaries
increases
with increasing mass ratio, or it might also be flat. Long period systems on
the other hand show the converse behaviour. At periastron it appears possible
that the secondary proto star accretes mass
from the larger circumstellar disk of the primary as noted by Bonnell \&
Bastien (1992). We adopt the following
simple model by assuming the mass ratio of the system,
$q=m_2/m_1\le1$, changes from a birth value to a final value given by

$$q_{\rm fin} = q_{\rm birth} + (1-q_{\rm birth})\,\rho^*,   \eqno(6)$$

\noindent where

$$\rho^* = \cases{\rho, &if $\rho\le1$; \cr
                1,     &if $\rho>1$.   \cr} \eqno(7)$$

\noindent The final mass of the secondary is given by $m_{\rm 2,fin}=q_{\rm
fin}m_{\rm 1,birth}$, where $m_{\rm 1,birth}$ is the birth mass of the
primary, which we assume does not change.
This implies a net gain in mass of the binary stellar system
which may be as large as $m_{\rm 1,birth}-m_{\rm 2,birth}$.
An alternative model based on equations~6 and~7 but constraining the total
mass of the binary system to remain constant proved to be unsatisfactory when
comparing with the short period observational data.

We stress that we consider the {\it feeding} hypothesis merely to propose that
a different formation mechanism for short period binaries need not be
necessary. We do not exclude this possibility but note that some sort of mass
exchange or accretion sharing in proto-stellar systems with $\rho\approx1$ must
be expected. We must however always bear in mind that the evidence
for a different distribution of mass ratios for long and short period systems
is weak because the number of data in the short period sample is small.

A reduction of eccentricity implies a reduction of orbital period and
we need to find  log$_{10}P_{\rm min}>0$ in equation~1 in
order to reproduce the observed main sequence period distribution.
Depending on the amount of pre-main sequence eigenevolution
we allow, we need a lower cutoff, log$_{10}P_{\rm min}$, in the log(period)
distribution
which increases with increasing proto-stellar orbital evolution. However,
log$_{10}P_{\rm min}<3$, because pre-main sequence binaries
with log$_{10}P_{\rm min}>3$
show no strong orbital evolution (figure~1 in K1). There
are not enough initial systems with log$_{10}P>3$ and sufficiently large
eccentricity to fill the $e-{\rm log}_{10}P$ diagram at log$_{10}P<3$. Also, in
paper K1 we showed that stimulated evolution does not suffice to harden enough
orbits. We thus expect log$_{10}P_{\rm min}<2$.

We choose in equation~1
${\rm log}_{10}P_{\rm min}=1$, $\eta=2.5$, $\delta=45$ and obtain

$$ f_{\rm P,birth} = 2.5 \, { \left({\rm log}_{10}P - 1 \right)
  \over 45 + \left({\rm log}_{10}P - 1 \right)^2},  \eqno (8)$$

\noindent so that log$_{10}P_{\rm max}=8.43$ from equation~11b in K1.

\nobreak\vskip 10pt\nobreak
\noindent{\bf 2.3 Simulations -- finding $\lambda$ and $\chi$}
\nobreak\vskip 10pt\nobreak
\noindent
We need to establish which values for
$\lambda$ and $\chi$ in equation~4 best approximate the observational data.
We proceed as follows: 600 birth
component masses are chosen from the KTG(1.3) mass
function.
Birth eccentricities are generated from equation~4 in K1 and
birth periods from equation~8. We have
300~systems with uncorrelated component masses and distribute these five times
giving in total
1500~systems.

Increasing $\lambda$ leads to an increasing
circularisation cutoff period, and decreasing $\chi$ leads to a flattening of
the
upper envelope in the $e-{\rm log}_{10}P$ diagram. By adjusting both parameters
we are able to model the observed upper envelope and the circularisation
period. We find $\lambda=28$ and $\chi=0.75$ are the best
approximations to the $e-{\rm log}_{10}P$ data of Duquennoy \& Mayor (1991).
This model is depicted in Fig.~2, and from the central and bottom panels we
observe that the shape of the eccentricity and period distributions for
$1\le{\rm log}_{10}P\le3$ are also in good agreement with the data.

\nobreak\vskip 10pt\nobreak
\noindent{\bf 2.4 The initial eccentricity distribution}
\nobreak\vskip 10pt\nobreak
\noindent In this section we demonstrate the very different character of
eigenevolution
and stimulated evolution and discuss the efficiency of thermalising an initial
eccentricity distribution in a dominant mode cluster.

In the top panel of Fig.~3 we visualise the change of eccentricity and
period owing to our model for eigenevolution
if $\lambda=28$ and $\chi=0.75$. We generate binary systems with
birth eccentricities $0.9, 0.5$ and~0.1 with periods from equation~8,
the range at birth of which are shown by the dotted lines.
{}From the figure we see that orbits with $e_{\rm birth}>0.5$
are circularised for $P<1\,$days approximately (for this illustrative example
we do not merge binaries with semi major axis less than $10\,R_\odot$).

In the lower panel of Fig.~3 we plot the distribution of birth
eccentricities (dotted lines) and the
distribution of
eccentricities after cluster dissolution for $N_{\rm run}=2$ simulations of a
cluster consisting initially of
200~binaries with $R_{0.5}=0.85\,$pc. We observe that hardening and softening
of binaries is a rare occurrence and
that stimulated evolution becomes significant (i.e. the distribution of
orbits in the diagram is affected noticeably) only at log$_{10}P>3$
approximately
in such a cluster. For $e_{\rm birth}=0.9,0.5,0.1$ we find that the final
proportion of binaries after the cluster has disintegrated is $f_{\rm
tot}=0.47,0.46,0.49$, respectively. There is no measurable
correlation of
the ionisation of binaries with initial eccentricity because the
binding energy of the binary is a function of orbital period only. The
resulting distributions of periods and mass ratios also do not change
significantly if different initial eccentricities are used.
We note however from Fig.~3 that after aggregate dissolution the maximum
period, $P^{\rm max}_{\rm orig ecc}$,
below which orbits with original eccentricity remain
decreases with increasing initial eccentricity: For $e_{\rm birth}=0.1, 0.5,
0.9$ we infer from
the figure that log$_{10}(P^{\rm max}_{\rm orig ecc})\approx7.4, 6.3, 5.9$,
respectively.
This suggests that circular orbits are more stable against stimulated
evolution, which is to be expected because stellar components
on an eccentric orbit spend a large time at large relative separation thus
being more susceptible to near-neighbour perturbation (see Aarseth 1992 for a
more in depth discussion). We expect to observe
this in the distribution of orbital elements and the proportion of final
binaries if a sufficiently large number of simulations of each
initial eccentricity is performed to reduce the statistical uncertainties
significantly, but clearly the effect is not very pronounced. Excluding
$e=e_{\rm birth}$, we find that
the resulting distributions after cluster disintegration
approach the thermal distribution, as shown in Fig.~4.

These simulations show
that an initially significantly different eccentricity distribution than the
thermal distribution
cannot dynamically relax to the thermal distribution (equation~4 in K1) in our
dominant mode cluster. If most stars are born in aggregates that are
dynamically equivalent to our dominant mode cluster then binary systems must
be born with a birth eccentricity distribution which is approximately
dynamically relaxed. This conclusion is also arrived at by Aarseth (1992).

\nobreak\vskip 10pt\nobreak
\noindent{\bf 2.5 `Birth' vs `initial' orbital parameter distributions}
\nobreak\vskip 10pt\nobreak
\noindent
We differentiate between the {\it birth} distributions and {\it initial}
distributions. The initial distributions result from the birth distributions
after pre-main sequence eigenevolution has largely completed.
The initial distributions are the
distributions of orbital elements of pre-main sequence stars after most of the
circum-proto stellar material has disappeared through accretion or expulsion.
With `initial' we refer to the time ($t=0$) at which the N-body integration
starts, which is equivalent to that `instant' in time when the age of the
system becomes sufficiently large that dynamical evolution
of the stellar cluster becomes effective.
Stimulated evolution then evolves the initial distribution to the
main sequence distribution we observe in the Galactic field. Thus, strictly
speaking, our birth distribution must be seen in context with our model for
pre-main sequence eigenevolution only.

The model data in Fig.~1 represent our
birth distributions, whereas the data in Fig.~2 are the
initial distributions that are established at an age of about $10^5$~yrs.

\bigskip
\bigbreak
\noindent{\bf 3 THE DOMINANT MODE IN STAR FORMATION?}
\nobreak\vskip 10pt\nobreak
\noindent
We show here that when our
{\it initial} orbital parameter distributions are subject to stimulated
evolution in the dominant mode cluster
they evolve to distributions which reproduce the observed
distributions for main sequence stars. We also discuss the formation of triple
and quadruple systems by capture.

We assume $N_{\rm bin}=200$ binary
systems are formed in the dominant mode cluster, as described in the
Introduction. The
half mass radius of $0.85\,$pc (table~1 in K1) is somewhat larger than
our previous
$R_{0.5}=0.77\,$pc (K1). We increase $R_{0.5}$ somewhat on the basis
of additional numerical experiments with our new period distribution by
monitoring depletion of the mass-ratio and period distribution, $f_{\rm q}$ and
$f_{\rm P}$, respectively. Birth eccentricities and
stellar masses are generated as in Section~3.2.1 of K1. Birth periods are given
by
equation~8. We perform~20
simulations to obtain useful statistics. In addition to the
eigenevolution of the orbital elements introduced in Section~2.2 we merge
components of a binary to a single star with combined mass of the components
if the semi-major axis after eigenevolution is $a\le10\,R_\odot$.

In our simulations stars are approximated by point particles. The equations of
motion
of the stars in a binary system are not subject to dissipative forces that
drive
the eigenevolution simultaneously to being subject to perturbative forces from
other systems in close proximity. We believe this approximation is permissible
on grounds of our argumentation in Section~2.1.

\nobreak\vskip 10pt\nobreak
\noindent{\bf 3.1 The eccentricity--period diagram}
\nobreak\vskip 10pt\nobreak
\noindent Evolution of
the cluster begins with the binary systems having eccentricity and
period distributions after pre-main sequence eigenevolution as
shown in the top panel of Fig.~5a. We have initially 4000
binary systems of which 124 merge so that at $t=0$, 3~per~cent of all
systems have merged to single stars. While we expect merging to occur during
pre-main sequence eigenevolution we caution against overinterpreting this
number given the simple nature of our eigenevolution model.

The intial distribution
of binaries (top panel of Fig.~5a) evolves before and during cluster
disintegration to the
$e-{\rm log}_{10}P$ distribution shown in the bottom panel of Fig.~5a. We
now find binaries in the eigenevolution region which have been placed there via
stimulated evolution. The number of binaries to the left of the constant
periastron curves (short dashed lines) with $e>0.3$ is small amounting to
between~1 and 2~per cent of all systems. We expect to observe
{\it forbidden orbital elements} in stellar clusters. Examples of such orbits
(see table~3 in Duquennoy et al. 1992) may be vB~164, vB~121 in the Hyades
cluster, KW~181 in the Praesepe cluster and S1284 in M67, all of which have
$e>0.2$ although their periods are significantly less than the cutoff period.
However, these binaries have $e<0.4$, whereas our forbidden orbits (Fig.~5a)
have $e>0.6$. This discrepancy is not serious because tidal circularisation
will begin reducing the eccentricity of a forbidden orbit immediately it has
been created so that our forbidden orbits will eigenevolve to smaller
eccentricities. For example, a forbidden orbit with $e>0.8, P=10^2$~days and
$m_{\rm tot}=0.64\,M_\odot$ has $R_{\rm peri}<16\,R_\odot$ so that
eigenevolution may proceed rapidly, even for main sequence stars. Assessment of
the time it takes for this to occur is, however, difficult, but may be less
than 500~Myrs (see the examples presented in table~3 in Duquennoy et al.
1992) which is sufficient since the Hyades and Praesepe Clusters are older
than this.

The short eigenevolution timescale for pre-main sequence binaries (Zahn \&
Bouchet 1989) may make chance detection of forbidden orbits unlikely in very
young clusters.
Examples of such systems may, however, be P2486 and EK~Cep, the orbital
elements for which are listed in table~A2 of the review by Mathieu (1994). Both
may
have been pushed into the eigenevolution region with larger
eccentricities than currently observed. P2486 is located
in the Trapezium Cluster whereas EK~Cep is isolated. It may have been ejected
within the last $10^5$~yrs from an embedded aggregate or a small group of
pre-main sequence stars. A proper motion and radial velocity measurement would
be illuminating.

It is important to emphasise that any discussion of tidal circularisation
theory
(e.g. see Mathieu et al. 1992) must bear in mind that stimulated evolution
rather than incomplete tidal circularisation or eccentricity driving from disks
may be responsible for eccentric orbits with periods shorter than the
circularisation cutoff period.
The number of forbidden orbits observed in Galactic
clusters is a function of the rate of production of such orbits through
encounters, and the rate at which such orbits are removed through
eigenevolution, as well as the rate at which systems are lost from
the cluster.

After cluster dissolution we have 2555 binaries and 2766 single stars (in
Section~3.4
we discuss how these are distributed in triple and quadruple systems) of which
124 are merged binaries. Thus, 4~per~cent of all single main sequence stars are
merged binaries according to the present model (but see caveat above).

Eigenevolution proceeds also for main sequence binaries with small period via
tidal dissipation but on a much longer time scale (see Mathieu et al. 1992 for
a discussion) and can in principle be simulated by a direct N-body integration
program (Aarseth 1994). However, this is not possible yet and we
model this with equations~2--5 but with different
$\lambda$ and $\chi$. Feeding
(equations~6 and~7) is not assumed to occur during main sequence
eigenevolution, and we do not merge components.  Since
we compare our model distribution with the observed main sequence distribution
(Duquennoy \& Mayor 1991) for which all orbits with a period less than
12~days are circularised we require $\lambda_{\rm ms}=24.7$ assuming a mean
mass of $1.4\,M_\odot$. Also, for small initial
eccentricities Zahn (1977) obtains $\chi_{\rm ms}=8$ ($\chi_{\rm ms}=5$ on the
other hand is
suggested by Goldman \& Mazeh 1991, but the difference is not important here).
The mean stellar age of the Duquennoy \& Mayor (1991) sample is about
$5\times10^9$~yrs which sets the timescale for main sequence eigenevolution. We
note that a circularisation cutoff period of 12.4~days is suggested by the
Galactic cluster M67 which is 4~Gyrs old (Mathieu 1994).

In Fig.~5b we show the $e-{\rm log}_{10}P$ diagram which results from the
distribution shown in the lower panel of Fig.~5a after main sequence
eigenevolution. The model main sequence
$e-$log$_{10}P$ distribution is now in good agreement with the
observational data. In particular the upper envelope is reproduced, and
orbits with approximately $P<11\,$days are circularised although the precise
cutoff period is mass dependent. The proportion of all
systems appearing in Fig.~5b with circularised orbits amounts to approximately
4~per~cent, and depends on log$_{10}P_{\rm min}$ (equation~8), eigenevolution
and merging during pre-main sequence eigenevolution.
Again we caution against overinterpreting this number, and stress that the
astrophysical details of the evolution of close binary systems
(tidal circularisation, Roche lobe overflow, merging) are quite beyond our
simple model and our present aim.

\nobreak\vskip 10pt\nobreak
\noindent{\bf 3.2 The main sequence orbital parameter distributions}
\nobreak\vskip 10pt\nobreak
\noindent
In Fig.~6 we compare our initial model eccentricity
distributions at $t=0$ (but after pre-main sequence eigenevolution) and after
cluster disintegration
(and after main sequence eigenevolution as in Fig.~5b) with main sequence
data for short-period and long period binaries. Agreement in both
cases is good.

The period
distributions initially (but after pre-main sequence eigenevolution) and after
cluster dissolution (after main sequence eigenevolution as in Fig.~5b) are
compared with the observational data in Fig.~7. Agreement is very
good. The final binary
proportion is $f_{\rm tot}=0.480\pm0.032$ which compares very
well with the observed $f_{\rm tot}^{\rm obs}=0.47\pm0.05$ in the Galactic
field (K1). We
also note that the initial period distribution (which is the pre-main sequence
eigenevolved birth distribution) is well approximated by the distribution given
by equation~1 (the 2$^{\rm nd}$ iteration in K1).

The observed mass
ratio distributions for G dwarf binaries (Duquennoy \& Mayor 1991) are
contrasted with our model distributions of secondary masses for primaries
with a mass in the range 0.9~to~1.1$\,M_\odot$ in Fig.~8. The top
panel shows that the short period model and observational distributions are in
acceptable agreement.
The very different character of this distribution contrasted with the
long period distribution shown in the bottom panel may be due to shared
accretion (i.e. feeding, Section~2.2).
Agreement with the observed main sequence distribution of secondary masses
for long-period systems is
excellent, and we care to remember that the initial distribution is the
power-law KTG(1.3) mass function.

\nobreak\vskip 10pt\nobreak
\noindent{\bf 3.3 Variation of binary proportion with system mass}
\nobreak\vskip 10pt\nobreak
\noindent
The variation of the proportion of binaries with the mass of the primary can be
used to constrain binary formation scenarios.
We define

$$  f_{\rm m} = {N_{\rm bin,m} \over N_{\rm sing,m} + N_{\rm bin,m} }, \eqno
(9)$$

\noindent where $N_{\rm bin,m}$ is the number of binaries with a primary mass
$m$, and $N_{\rm sing,m}$ is the number of single stars with mass $m$.

Fischer
\& Marcy (1992) argue that the proportion of M~dwarfs that are binary ($f_{\rm
M}=0.42\pm0.09$) is smaller than the proportion of G~dwarfs that are binary
($f_{\rm G}\approx0.57$) and suggest this is
consistent with capture into potential
wells of two different depths. Kroupa et al. (1993), on the other
hand, show that variation of $f_{\rm m}$can be understood by random pairing of
stellar masses if $f_{\rm tot}\approx0.6$. Leinert et al.
(1993) find that the proportion of pre-main sequence binaries does not vary
with K-band magnitude and, on the basis of the results of Kroupa et al.
(1993), interpret this finding as evidence for $f_{\rm tot}(t=0)=1$.
We emphasise, however, that the observational data shown in Fig.~9 {\it do not}
necessarily imply a decreasing binary proportion with decreasing primary mass!

In Fig.~9 we show how $f_{\rm m}$ varies with primary mass if stars form in
binary systems with uncorrelated masses. At $t=0$ (after pre-main sequence
eigenevolution) we have
$f_{\rm m}\approx1$.
If the majority of stars in the Galactic field are born in aggregates
that are dynamically equivalent to the dominant mode
cluster, as suggested by K1, then $f_{\rm m}\approx0.5$ on the main sequence
(Fig.~9). The small increase in $f_{\rm m}$ from $m=1\,M_\odot$ to
$m=0.4\,M_\odot$ comes about because the highest mass stars (the G~dwarfs in
our simulations) spend more time near the cluster centre than the less massive
stars. We expect $f_{\rm m}$ to flatten or decay somewhat if more realistic
embedded clusters
are modelled with no upper mass cutoff and a changing background potential.
The decrease to $f_{\rm m}=0.42$ at $m=0.2\,M_\odot$ results
because these low mass binaries are, on average, less bound than the higher
mass binaries.

\nobreak\vskip 10pt\nobreak
\noindent{\bf 3.4 Triple and quadruple systems}
\nobreak\vskip 10pt\nobreak
\noindent
On the main sequence Duquennoy \& Mayor (1991) observe the following ratio of
single~: ~double~: ~triple~: ~quadruple G-dwarf systems: 1.50~: ~1~: ~0.105~:
{}~0.026, although they emphasise that the number of triple and quadruple
systems
may be larger.

For the above ratios we have in our model 0~: ~1~: ~0~: ~0 at birth. We expect
that
some triple and quadruple systems form by capture at later times. We search the
output data of our 20 simulations for such systems by replacing all binary
systems by centre of mass particles, and searching for all binary centre of
mass systems using the same algorithm as described in K1. We identify all those
systems which have a binding energy between the centre of masses
$-E_{\rm bin}<0$ and which persist for at least 43~Myrs at $t=1\,$Gyr (i.e. the
same system must appear in two subsequent output lists). This is necessary
because transient binary centre of mass systems with very small binding energy
do appear. We find at $t=1$~Gyr 37 binary centre of mass systems.

Our analysis identifies 2743 single stars, 2504 binary, 23 triple and 14
quadruple systems with relative proportions 1.10~:~1~:~0.0092~:~0.0056.
Comparison
with the main sequence proportions demonstrates that formation by capture
cannot account for the number of triple and quadruple systems in the Galactic
field.

In Fig.10 we show the eccentricity--period diagram of the outer orbit of the
triple and quadruple systems formed by capture. We refer to `outer' orbital
parameters as the orbital parameters of the binary which consists of two centre
of mass particles, each of which can either be a single star or an `inner'
binary. All orbits of triple systems lie in the range
$10^7<P_{\rm outer}<10^{10.2}$~d, whereas the quadruple systems have
$10^{8.2}<P_{\rm outer}<10^{11}$~d. The upper limit is given approximately by
the Galactic
tidal field (K1) and increases with system mass. We observe from
Fig.~10 that quadruple systems
have outer eccentricities $e_{\rm outer}<0.8$ approximately, while
eccentricities close to unity are
quite common in triple systems. The outer mass ratio distribution is biased
towards one in quadruple systems (by the nature of our stellar mass range)
whereas triple systems show a peak at $q_{\rm outer}\approx 0.3$ (Fig.~10). The
distribution of the ratio of inner to outer semi-major axes is bell shaped for
both triple and quadruple systems. As apparent from Fig.~10 both have a broad
maximum at log$_{10}\left(a_{\rm inner}/a_{\rm outer}\right)\approx-4$.
However, there are no quadruple systems with
log$_{10}\left(a_{\rm inner}/a_{\rm outer}\right)<-2.5$ (see also Fig.~11). The
distribution of orbits in the $e_{\rm outer}$--log$_{10}\left(a_{\rm
inner}/a_{\rm outer}\right)$ diagram is shown in Fig.~11, where we plot two
log$_{10}\left(a_{\rm inner}/a_{\rm outer}\right)$ values for one quadruple
system outer eccentricity. The distribution of eccentricities is approximately
thermal for triple systems, but $e_{\rm outer}>0.8$ does not occur in quadruple
systems
owing to the larger interaction cross section of the two inner binaries.

The bound nature and origin of the Proxima~Cen--$\alpha$~Cen~A/B triple system
remains unclear (Kamper \& Wesselink 1978, Matthews \& Gilmore 1993, Anasova,
Orlov \& Pavlova 1994). Proxima~Cen has a mass of $m_{\rm
C}=0.12\,M_\odot$ (based on $M_{\rm V}=15.49$ and the mass--$M_{\rm V}$
relation tabulated by Kroupa et al. 1993), the mass of
$\alpha$~Cen~A is $m_{\rm A}=1.10\,M_\odot$ and of
$\alpha$~Cen~B it is $m_{\rm B}=0.91\,M_\odot$.
The inner binary $\alpha$~Cen has $a_{\rm inner}=23.4\,$AU
and $e_{\rm inner}\approx0.5$. The current separation of Proxima from the inner
binary is approximately 13000~AU, i.e. 0.063~pc. Two systems in our model
Galactic field data resemble this triple system:
(i) $m_{\rm C}=0.39\,M_\odot, m_{\rm
A+B}=1.72\,M_\odot, e_{\rm inner}=0.40, a_{\rm inner}=33\,$AU with a separation
at $t=1$~Gyr of 7685~AU. The outer orbit has $a_{\rm outer}=6419$~AU, $e_{\rm
outer}=0.92$ and log$_{10}P_{\rm outer}=8.1$;
(ii) $m_{\rm C}=0.13\,M_\odot, m_{\rm
A+B}=1.53\,M_\odot, e_{\rm inner}=0.44, a_{\rm inner}=61\,$AU with a separation
at $t=1$~Gyr of 68858~AU. The outer orbit has $a_{\rm outer}=46985$~AU,
$e_{\rm outer}=0.51$ and log$_{10}P_{\rm outer}=9.5$.

To obtain some bounds on the likely outer orbital parameters of the
Prox~Cen--$\alpha$~Cen~A/B system we note from the upper panel of Fig.~10 that
log$_{10}P_{\rm outer}<10.2$. Kepler's equation and the total mass of the
system ($2.1\,M_\odot$) gives $a_{\rm outer}<1.58\times10^5$~AU, i.e.
log$_{10}(a_{\rm inner}/a_{\rm outer})=\kappa>-3.8$. More generally, assume
$R_{\rm peri,outer}=10^\theta a_{\rm inner}$. Then $e_{\rm
outer}<1-10^{\theta+\kappa}$.
Since $\kappa=-1.5$ implies $e_{\rm outer}=0$, approximately (top panel of
Fig.11),
we obtain $\theta=1.5$. Thus our numerical experiment implies that triple
systems formed by capture
are stable if $R_{\rm peri,outer}>10^{1.5} a_{\rm inner}$,
$a_{\rm outer}>10^{1.5} a_{\rm inner}$, and $e_{\rm outer}<1-10^{1.5+\kappa}$.
In the upper panel of Fig.~11 we indicate the approximate region region
which is allowed for Proxima~Cen--$\alpha$~Cen~A/B.

A detailed study of the stability of triple systems initially on circular
orbits is provided by Kiseleva, Eggleton \& Orlov (1994). From their table~1 we
find that the Proxima--$\alpha$~Cen~A/B system can be stable only if $P_{\rm
outer}/P_{\rm inner}>4$ approximately, i.e. $\kappa<-0.4$.
This represents a strict upper limit on $\kappa$, and we note that if
$a_{\rm outer}\approx13000$~AU then $\kappa\approx-2.7$.

The capture scenario for the Proxima~Cen--$\alpha$~Cen~A/B system presented
here
does not alleviate the problem posed by the apparent youth of Proxima and the
advanced age of $\alpha$~Cen~A/B, which may be older than the Sun (Matthews \&
Gilmore 1993), unless either flare
activity in late type M~dwarfs persists for about $5\times10^9$~yrs (see
also the discussion in section~10.3 in Kroupa et al. 1993) or
Proxima was captured in the Galactic field by $\alpha$~Cen~A/B which is
highly improbable.

However, we cannot exclude the
possibility that this triple system had initially a much larger $\kappa$ and
that we are currently observing Proxima being ejected, or nearly ejected.
Although this is very unlikely, this hypothesis may provide an explanation for
the persisting significant flare activity of Proxima (Kroupa, Burnam \& Blair
1989).
Relatively recent close passages of Proxima past $\alpha$~Cen~A/B may have
temporarily induced flare activity through tidal deformation.

\nobreak\vskip 10pt\nobreak
\noindent{\bf 3.5 The mass ratio distribution}
\nobreak\vskip 10pt\nobreak
\noindent
In the top panel of Fig.~12 we show as the
dashed histogram the initially uncorrelated mass-ratio distribution of all
binaries. The minor peak at
$q\approx0.9$ and the systems with $q=1.0$ result from our feeding hypothesis
(equations~6 and~7).
We find that after the cluster has dissolved about 70 per cent binaries survive
with $q\approx0.9$, whereas only 40 per cent survive with $q\approx0.2$. The
initial and final mass-ratio distributions are tabulated in Table~1.

\bigbreak
\vskip 3mm
\bigbreak

\hang{ {\bf Table 1.} The Mass-Ratio Distribution }

\nobreak
\vskip 1mm
\nobreak
{\hsize 15 cm \settabs 7 \columns

\+&$q$ &$N_q$ &$\delta N_q$ &$N_q$ &$\delta N_q$\cr
\+&    &$t=0$ & &final\cr
\+\cr
\+&0.05 & 0.0 &0.0 & 0.2 &0.1 \cr
\+&0.15 & 7.0 &0.6 & 2.7 &0.4 \cr
\+&0.25 &17.3 &1.0 & 8.2 &0.7 \cr
\+&0.35 &21.1 &1.1 &12.4 &0.8 \cr
\+&0.45 &23.0 &1.1 &15.1 &0.9 \cr
\+&0.55 &22.0 &1.1 &15.2 &0.9 \cr
\+&0.65 &22.4 &1.1 &14.3 &0.9 \cr
\+&0.75 &20.6 &1.0 &15.7 &0.9 \cr
\+&0.85 &26.2 &1.2 &17.5 &1.0 \cr
\+&0.95 &26.4 &1.2 &19.7 &1.0 \cr
\+&1.05 & 8.1 &0.7 & 7.7 &0.6 \cr
\+&1.15 & 0.0 &0.0 & 0.0 &0.0 \cr
}
\bigbreak\vskip 3mm

\hang{Notes to table:

$q=m_2/m_1\le1$, $m_1\le1.1\,M_\odot$ and $m_2\le1.1\,M_\odot$ are the
mass of the primary and secondary, respectively.

$N_q$ is the average number of systems in mass ratio bin $q_i$, $N_q={1\over
N_{\rm run}}\sum^{N_{\rm run}}_{i=1} \, N_{q_i}$.

$\delta N_q$ is the standard deviation of the mean,
$\delta N_q = \left[ {\sum^{N_{\rm run}}_{i=1}  \, N_{q_i}
                      \over (N_{\rm run}-1)} \right]^{1\over2}\,
N_{\rm run}^{-{1\over2}}$,

where $N_{q_i}$ is the number of systems in bin
$q_i$ in simulation $i$.

These model mass ratio distributions result from our uncorrelated birth
distribution and the KTG(1.3) mass function after pre-main sequence
eigenevolution (columns 2 and 3) and after stimulated evolution in the dominant
mode cluster (columns 4 and 5). The data tabulated here are averages of $N_{\rm
run}=20$ simulations.

The binaries appearing in the $q=1.05$ bin have $q=1$ and result from
our feeding hypothesis (Section~2.2).
}

\bigbreak
\vskip 3mm
\bigbreak

In
the bottom panel of Fig.~12 we show the stellar mass functions of the primaries
and secondaries initially (after pre-main sequence eigenevolution)
and after cluster disintegration. The mass function of the secondaries is
steeper than that of the primaries.
In Fig.~12 both the primary and secondary star mass functions show about the
same depletion except at
$m_2>0.5\,M_\odot$, where the secondary star mass function is more depleted
after
the cluster has disintegrated. Mass segregation implies that the massive
binaries
spend more time in a dense environment so that the systems with massive
components are ionised to a greater extent than systems with low mass
components (see also Fig.~9). Depletion of orbits as a function of $q$ is
not a process that can be modelled straightforwardly (see also the discussion
of equation~6 in K1) and depends on the lifetime of the cluster and its initial
configuration.

\bigskip
\bigbreak
\noindent{\bf 4 THE INITIAL ANGULAR MOMENTUM DISTRIBUTION}
\nobreak\vskip 10pt\nobreak
\noindent The initial
distributions of orbital elements are equivalent to an initial
distribution of angular momenta of the
binary systems. The angular momentum is given by (masses in $M_\odot$, $P$ in
days)

$$J = 1.24\times10^{52}\,
(1-e^2)^{1\over2}\,P^{1\over3}\,{m_1\,m_2\over\left(m_1+m_2\right)^{1\over3}}\,
      {\rm~~~~~}\left({{\rm g}\,{\rm cm}^2\over {\rm sec}}\right). \eqno (10)$$

The distributions of specific angular momenta,
$J/M=J/(m_1+m_2)$, are shown in Fig.~13 at $t=0$
(pre-main sequence
eigenevolution has a negligible effect on the specific angular momentum
distribution) and after cluster dissolution. Ionisation of
the least bound binaries leads to a depletion of the specific angular momentum
distribution at high specific angular momenta.
Our model initial binary star distribution
adjoins the molecular cloud distribution without a gap (compare with
Simon 1992).
The molecular cloud core data are incomplete at small $J/M$ owing to the
observational difficulties in detecting slowly rotating cores (Goodman et al.
1993) and the distribution may extend to much lower values. An in depth
discussion of angular momentum redistribution during star formation can be
found in Bodenheimer et al. (1993).

\bigskip
\bigbreak
\noindent{\bf 5 IMPLICATIONS FOR CIRCUMSTELLAR DISKS}
\nobreak\vskip 10pt\nobreak
\noindent
The orbital period of Jupiter is log$_{10}P=3.64$ and of Pluto it is
log$_{10}P=4.96$. The Sun may have had a companion with log$_{10}P\ge6$
permitting
stability of the solar system. From Fig.~7 we find that initially
approximately 44~per~cent of all G~dwarfs were in such binary systems. These
consist predominantly of one G~dwarf and a late-type companion, and rarely two
G~dwarfs. After cluster
disintegration about 12~per~cent of all G~dwarfs remain in such binary systems.
Of all G~dwarfs $44-12=32$~per~cent end up as
single stars that came from binary systems with log$_{10}P\ge6$. The majority
of the ex-companions of these are K- and M-dwarfs.
Since the components from ionised binary systems initially with
log$_{10}P\ge6$ are are not likely to be ejected from the aggregate
immediately we need to estimate
the proportion of systems which have encounters that are destructive for
log$_{10}P\le5$ to estimate the proportion of components that may retain a
possible planetary system.
{}From Fig.~7 we find that 9~per~cent of all systems with log$_{10}P\le5$
are ionised. Thus, $0.91\times0.32=0.29$, i.e. 29~per~cent of all G~dwarfs are
single and
are likely to have escaped additional close encounters that are destructive at
log$_{10}P\le5$.

While reality will be much more
complicated than discussed here, we have shown that in principle
our Sun may be one such single G~dwarf.
Systems with
log$_{10}P\ge6$ may also have a solar-type planetary system. We may
thus expect in total $29+12=41$~per~cent of all G~dwarfs may harbour solar-type
planetary systems.

The total angular momentum vector of the solar planetary system is tilted
by~7$^{\rm o}$ with repsect to the rotation axis of the Sun, and the
orbit of Pluto is inclined by $17^{\rm o}$ to the ecliptic. This
may indicate an early perturbation of the outer reaches of the
solar system by either a young companion or a ``visit'' by a neighbour in the
stellar group in which the sun was born. Heller (1993) models the perturbations
of circumstellar disks in stellar groups and finds that this interpretation of
the tilt of the total angular momentum vector of our planetary system is
reasonable.

We can extend these considerations to circumstellar disks around all late-type
stars. From fig.~1 in K1 or Fig.~7 we note that $f_{\rm P}$ does not depend on
spectral
type. Two stars result from each ionisation event. If the present model of
star-formation is a valid approximation to reality then about 40~per
cent of all late-type stars we see on the sky are single stars which at birth
had a companion in an orbit with log$_{10}P\ge6$. These stars may have disks
(when younger than a few Myrs) or planetary systems that extend to at least
40~AU.

\bigskip
\bigbreak
\noindent{\bf 6 CONCLUSIONS}
\nobreak\vskip 10pt\nobreak
\noindent
In K1 we show that stimulated evolution in a stellar
aggregate consisting of a few hundred binaries and with a half mass radius
between 0.08~pc and 2.5~pc cannot evolve a period distribution which is
initially limited
to $3\le{\rm log}_{10}P\le 7.5$ into a distribution with a sufficient number of
orbits at log$_{10}P<3$ to account for the observational data. The observed
distribution of periods can only be accounted for if binary systems appear at
the birth line
with periods ranging from a few days to $10^9$~days. We approximate the {\it
initial} period distribution by equation~1. The
resulting initial distribution of
specific angular momenta for binary systems (Section~4) adjoins the specific
angular momentum distribution of molecular cloud cores (Fig.~13).

We also assume that, rather than reflecting a
different star-forming mode for short-period systems, the
difference between the short and long period mass
ratio and eccentricity distributions are manifestations of orbital element
evolution which affects systems with small periastron (Section~2).
We assume this mechanism is
intrinsic to the young binary system and occurs to near-completion during the
proto stellar accretion phase. We
develop a simple model of {\it eigenevolution} (Section~2.2) according to which
the
eccentricity--period diagram is depopulated at large eccentricities and small
periods. We term this the {\it eigenevolution region} of the diagram and refer
to any orbits within this region as {\it forbidden orbital parameters}.
Thus the eigenevolution region is assumed to depopulate
on a timescale of about 0.1~Myr (for pre-main sequence stars), establishing
the
upper envelope of the initial distribution of orbits in the \e-p diagram by the
time the pre-main sequence binaries appear near the birth-line in the HR
diagram. Stimulated
evolution depletes the distribution of orbits with approximately
log$_{10}P>4$ on the dynamical evolution timescale of the stellar cluster.

Correcting for pre-main sequence eigenevolution and stimulated evolution we
derive a {\it birth} distribution of periods (Section~2.5, equation~8).

To study in detail the evolution of the initial orbital parameters
we perform 20~simulations of the dominant mode cluster (Section~3) and are
left with a model Galactic field stellar population after its disintegration.
We find that
thermalisation of orbits in the dominant mode cluster
is not complete enough to allow other than an initially approximately
thermal eccentricity distribution (Section~2.4). The final
binary star population has a period distribution in good agreement with the
observed distribution of periods of G-, K- and M-dwarf
binaries (Fig.~7) whereby the rise of the log$_{10}P$ distribution
to log$_{10}P\approx4.8$ approximately reflects the initial distribution. The
decay of the distribution at log$_{10}P>4.8$ results from stimulated evolution.
The final short and long
period mass ratio and eccentricity distributions are in good agreement with
observational data for G~dwarfs (Figs.~8 and~6, respectively). In particular,
the maximum in the observed long-period
mass-ratio distribution of G-dwarf binaries at a mass ratio of about 0.25 is
reproduced even though the underlying KTG(1.3) mass function rises
monotonically
with decreasing stellar mass. We tabulate the overall model mass-ratio
distribution for Galactic field systems with component masses less massive than
$1.1\,M_\odot$ in Table~1.
The proportion of binaries in our model, $f_{\rm
tot}=0.48\pm0.03$, compares very well with the observed proportion $f_{\rm
tot}^{\rm obs}=0.47\pm0.05$.

The proportion of systems as a function of the mass of the primary star,
$f_{\rm m}$, in our model Galactic field population is in good agreement
with the observational constraints (Fig.~9). We emphasise that the latter are
consistent with no decrease of $f_{\rm m}$ with decreasing primary mass.

When comparing our model with the observational data we
keep in mind that in the observational period distribution triple
systems are counted as two orbits and
quadruple systems as three orbits adding data at the long-period end of the
observational period distribution. Our model does not extend to higher
multiplicities at birth.
The number of triple and quadruple systems formed by capture in our model is
too small to account for the observed numbers
(Section~3.4). Proxima~Cen--$\alpha$~Cen~A/B has orbital
parameters similar to triple systems formed in our simulations by capture in
the dominant mode cluster.

Stimulated evolution pushes about 1--2 per cent
of all final orbits into the eigenevolution region so
that we expect to find in clusters binaries which have eccentric orbits
although their period is shorter than the circularisation cutoff period.
The pre-main sequence binaries P2486 (in the Trapezium Cluster) and EK~Cep
(isolated) may be examples of such systems with forbidden orbital elements.
This interpretation of EK Cep requires it to have had a close encounter with
another system approximately within the last $10^5$~yrs and as a result of this
it may have a relatively large velocity.

The existence of our planetary system does not
contradict the hypothesis that all stars form in multiple systems
(Section~5), nor that most stars originate in clustered star formation.
We expect about 40~per cent of all young late-type stars to be single and have
disks extending to about 40~AU. The majority of the remaining stars have
companions.

We have demonstrated that reasonable assumptions about star formation that are
consistent with observations of young stars, and the assumption that most stars
form in aggregates that are dynamically equivalent to our
dominant mode cluster (see K1), lead to a model population of
stellar systems in the Galactic field which is in good agreement with the
available observational constraints.
Future observations (high-resolution observations of very young stars,
large-scale near-infrared imaging surveys of star forming regions, radial
velocity and proper motion surveys [see K3]) will verify if the
assumptions made in this paper are true. These are: (i) most stars form as
binaries,
(ii) these
have uncorrelated component masses, (iii) a distribution of orbits similar
to equation~1 (equation~8 is our approximation to the birth distribution and
is unlikely to be ever observable) and (iv) they form in aggregates dynamically
equivalent to our dominant mode cluster.

Our assumptions and results would appear to be valid if most proto-stellar
binary systems would form by dissipative capture in small groups of a few to
ten accreting proto-stars that are clustered in aggregates of tens of such
groups (K1).

\bigskip
\bigbreak
\par\noindent{\bf ACKNOWLEDGMENTS}
\nobreak
\nobreak
\noindent I am very grateful to Sverre Aarseth without whose
kind
help with and distribution of nbody5 this work would not have been
possible and for useful comments.
I am also grateful to Cathy Clarke for very useful discussions concerning the
\e-p diagram
and to Peter Schwekendiek for his rapid acquaintance with and
maintenance of the new computing facilities installed at ARI towards the end of
1993.

\bigskip
\noindent{\bf REFERENCES}
\nobreak
\bigskip
\nex Aarseth, S. J., 1992, N-Body Simulations of Primordial Binaries and Tidal
     Capture in Open Clusters. In:
     Duquennoy, A., Mayor, M. (eds), Binaries as Tracers of Stellar
     Formation, Cambridge Univ. Press, Cambridge, p.6
\nex Aarseth, S. J., 1994, Direct Methods for N-Body Simulations. In:
      Contopoulos, G., Spyrou, N. K., Vlahos, L. (eds.), Galactic Dynamics and
      N-Body Simulations, Springer, Berlin, p.277
\nex Anasova, J., Orlov, V. V., Pavlova, N. A., 1994, A\&A 292, 115
\nex Bodenheimer, P., Ruzmaikina, T., Mathieu, R. D., 1993, Stellar Multiple
     Systems: Constraints on the Mechanisms of Origin, In: Levy, E. H.,
     Lunine,
     J. I., Matthews, M.S. (eds), Protostars and Planets III, Univ. of Arizona
     Press, Tucson,
     p.367
\nex Bonnell, I., 1994, Fragmentation and the Formation of Binary and Multiple
     Systems, In: Clemens, D., Barvainis, R. (eds), Clouds, Cores, and Low Mass
     Stars, ASP Conf. Series, in press
\nex Bonnell, I., Bastien, P. 1992, The Formation of Non-Equal Mass Binary and
     Multiple Systems. In:
     McAlister, H. A., Hartkopf, W. I. (eds.) Complementary Approaches to
     Double and Multiple Star Research, Proc. IAU Coll. 135, ASP Conference
     Series, 32, p.206
\nex Duquennoy, A., Mayor, M., 1991, A\&A 248, 485
\nex Duquennoy, A., Mayor, M., Mermilliod J.-C., 1992, Distribution and
     Evolution of Orbital
     Elements for $1\,M_\odot$ Primaries. In:
     Duquennoy, A., Mayor, M. (eds), Binaries as Tracers of Stellar
     Formation, Cambridge Univ. Press, Cambridge, p.52
\nex Fischer, D. A., Marcy, G. W., 1992, ApJ 396, 178
\nex Goldman, I., Mazeh, T., 1991, ApJ 376, 260
\nex Goldman, I., Mazeh, T., 1994, ApJ 429, 362
\nex Goodman, A. A., Benson, P. J., Fuller, G. A., Myers, P. C., 1993, ApJ 406,
     528
\nex Heller, C. H., 1993, ApJ 408, 337
\nex Kamper, K. W., Wesselink, A. J., 1978, AJ 83, 1653
\nex Kiseleva, L. G., Eggleton, P. P., Orlov, V. V., 1994, MNRAS 270, 936
\nex Kroupa, P., 1995a, Inverse Dynamical Population Synthesis and Star
     Formation, MNRAS, in press (K1)
\nex Kroupa, P., 1995b, Star Cluster Evolution, Dynamical Age Estimation and
     the Kinematical Signature of Star
     Formation, MNRAS, in press (K3)
\nex Kroupa, P., Burnam, R. R., Blair, D. G., 1989, Proc. Astron. Soc.
     Australia 8, 119
\nex Kroupa, P., Tout, C. A., Gilmore, G., 1993, MNRAS 262, 545
\nex Leinert, Ch., Zinnecker, H., Weitzel, N., Christou, J., Ridgway, S. T.,
     Jameson, R., Haas, M., Lenzen, R., 1993, A\&A 278, 129
\nex Lubow, S. H., Artymowicz, P., 1992, Eccentricity Evolution of a Binary
     Embedded in a disk. In:
     Duquennoy, A., Mayor, M. (eds), Binaries as Tracers of Stellar
     Formation, Cambridge Univ. Press, Cambridge, p.145
\nex Matthews, R., Gilmore, G., 1993, MNRAS 261, L5
\nex Mathieu, R. D., Latham, D. W., Mazeh, T., Duquennoy, A., Mayor, M.,
     Mermilliod, J.-C., 1992, The Distribution of Cutoff Periods with Age. In:
     Duquennoy, A., Mayor, M. (eds), Binaries as Tracers of Stellar
     Formation, Cambridge Univ. Press, Cambridge, p.278
\nex Mathieu, R. D., 1994, ARA\&A 32, 465
\nex Mazeh, T., Goldberg, D., Duquennoy, A., Mayor, M., 1992, ApJ 401, 265
\nex Simon, M., 1992, Multiplicity Among the Young Stars. In:
     McAlister, H. A., Hartkopf, W. I. (eds.) Complementary Approaches to
     Double and Multiple Star Research, Proc. IAU Coll. 135, ASP Conference
     Series, 32, p.41
\nex Zahn, J.-P., 1977, A\&A 57, 383
\nex Zahn, J.-P., 1992, Present State of the Tidal Theory. In:
     Duquennoy, A., Mayor, M. (eds), Binaries as Tracers of Stellar
     Formation, Cambridge Univ. Press, Cambridge, p.253
\nex Zahn, J.-P., Bouchet, L., 1989, A\&A 223, 112

%==============FIGURES:========================
\vfill\eject

\centerline{\bf Figure captions}
\smallskip
\noindent {\bf Figure 1.}
{\bf Top panel}: The {\it birth} distribution of model data in the
eccentricity-period plane (Section~2.1). The thick
dashed line is the approximate upper envelope for
main sequence solar type binary systems (Duquennoy \& Mayor 1991). {\bf Middle
panel}: the birth eccentricity distribution for long (log$_{10}P>3$, solid
histogram)
and short ($1<$log$_{10}P<3$, dashed histogram) period binaries. The filled and
open circles show the corresponding observed G-dwarf distributions normalised
to unit
area as in fig.~2 in paper K1. {\bf Bottom panel}: The birth distribution of
periods
is shown as the thick continuous line (equation~8) and the solid histogram.
The dashed lines are the first and second iterations (fig.~8 in K1).
Pre-main sequence eigenevolution is not implemented here.
Observational data are as in fig.~1 of K1 (solid cirlces: G~dwarfs; open
squares and triangles: pre-main sequence binaries).

\vskip 5mm

\noindent {\bf Figure 2.} As Fig.~1.
The {\it initial} distribution of model orbital parameters obtained
after applying a pre-main sequence eigenevolution model with
$\chi=0.75$ and $\lambda=28$ to the data shown in Fig.~1 (Section~2.3). In
the {\bf top panel} the short dashed lines are
constant periastron loci (equation~2 with $R_{\rm peri}=\lambda\,R_\odot$) for
system masses of $m_{\rm tot}=2.2,0.64,0.2\,M_\odot$ in increasing thickness.
Systems lying to the left of these lines have periastron distances less than
$28\,R_\odot$. Binaries with semi major axis less than $10\,R_\odot$ are not
merged here. Note that in the {\bf bottom panel} the solid histogram represents
the
initial period distribution, whereas the solid curve is the birth distribution
(equation~8). See also discussion in Section~2.5. The observational
data are as in Fig.~1.

\vskip 5mm

\noindent {\bf Figure 3.} Eigenevolution and stimulated evolution of orbital
parameters (Section~2.4). {\bf Upper panel}: the effects of our model for
pre-main sequence eigenevolution
on birth eccentricities $e_{\rm birth}=0.1,0.5,0.9$, shown
by the horizontal dotted lines, is presented for $\chi=0.75$ and $\lambda=28$.
The thick long dashed line is as in Fig.~1.
The short dashed lines are as in Fig.~2. {\bf Lower panel}: Assuming no
eigenevolution we show the distributions after cluster disintegration that
result from stimulated evolution in the dominant mode cluster
for birth eccentricities as in the top panel. Histograms are shown in Fig.~4.
The results of two simulations ($N_{\rm run}=2$) are shown
for each birth eccentricity.
The birth period distribution in both panels is given by equation~8.

\vskip 5mm

\noindent {\bf Figure 4.} (Section~2.4) The eccentricity distributions after
cluster dissolution
are shown for the birth eccentricities ($e_{\rm birth}=0.1,0.5,0.9$) which are
plotted in  the lower panel of Fig.~3.
The histograms, represented by the same symbols as in Fig.~3, have been
normalised to unit area after discarding the remaining peaks at the original
eccentricities. The mean of two simulations are shown per birth eccentricity.
The solid line histogram is the observed distribution of eccentricities
for main sequence G dwarfs with log$_{10}P>3$ and normalised to unit
area (Duquennoy \& Mayor 1991).

\vskip 5mm

\noindent {\bf Figure 5.} (Section~3.1) {\bf a}: The $e-{\rm log}_{10}P$
diagram for our model
of the pre-main sequence eigenevolution ($\lambda=28,\chi=0.75$) at $t=0$
({\bf top panel}) and after cluster disintegration
({\bf bottom panel}) when 1164 bound binaries appear in
the plot. Systems with semimajor axis after pre-main sequence
eigenevolution $a\le10\,R_\odot$ have been merged. Stimulated evolution in
the dominant mode cluster
populates the eigenevolution region implying that {\it forbidden orbital
parameters} should be observable in stellar clusters.
{\bf b}: Eigenevolution on the main sequence with
$\lambda_{\rm ms}=24.7,\chi_{\rm ms}=8$ applied to the data in the bottom panel
of Fig.~5a depopulates the eigenevolution region and circularises all
orbits
with period less than about 11~days on a timescale of about 5~Gyrs. The thick
long-dashed and the short
dashed curves in Fig.~5a and~b are as in Fig.~2. We note the sparsity of data
with log$_{10}P>2$ with small eccentricity does not imply eccentricity driving
(Section~2.1). This figure contains data of 9 of the 20 simulations
(table~1 in K1).

\vskip 5mm

\noindent {\bf Figure 6.} (Section~3.2) {\bf Top panel}: The short period
eccentricity distribution for all simulated data, a subset of which is shown in
Fig.~5. The dashed line represents the intial ($t=0$) distribution
after pre-main sequence eigenevolution
and the solid histogram represents the final distribution
after application of main sequence eigenevolution (as in Fig.~5b).
The model distributions are averaged
from 20~simulations (table~1 in K1). Open
circles are the data for main sequence G dwarf binaries (Duquennoy \& Mayor
1991). {\bf Lower panel}: The same as the upper panel but for long period
systems.
The solid dots are main sequence G dwarf binay system data from Duquennoy \&
Mayor (1991). All distributions are normalised to unit area.

\vskip 5mm

\noindent {\bf Figure 7.} (Section~3.2) The main sequence period distribution
for all simulated data, a subset of which is shown in Fig.~5b,
is presented here as the solid histogram.
This distribution of periods results after the birth distribution (solid
curve, equation~8) is subjected to pre-main sequence eigenevolution, leading
to the initial distribution shown as the long-dashed histogram, and then to
stimulated evolution in the dominant mode cluster and main sequence
eigenevolution (the latter having
a negligible effect). The model distributions are averaged from 20~simulations
(table~1 in K1).
The short dashed curves are the first and second iteration (fig.~8 in K1). The
data are as in fig.~1 in K1 (solid circles: G~dwarfs; open circles: K~dwarfs;
stars: M~dwarfs; open squares and triangles: pre-main sequence binaries).

\vskip 5mm

\noindent {\bf Figure 8.} The G~dwarf mass-ratio distribution. {\bf Top panel}:
short period distribution of secondary star
masses in systems with primaries with a mass between 0.9~and 1.1$\,M_\odot$
(Section~3.2).
The dashed histogram
represents the initial distribution ($t=0$) after pre-main sequence
eigenevolution, and the solid histogram
is the final distribution. The model distributions are averaged
from 20~simulations (table~1 in K1) and have been normalised to the data (open
circles) at $q>0.4$. The open circles
are G dwarf main sequence short period binary star data (Mazeh et al.
1992 and fig.~2 in K1). {\bf Bottom panel}: the same as top panel but
for long period systems. The solid dots are G~dwarf main sequence long period
binary star data (Duquennoy \& Mayor 1991 and fig.~2 in K1).

\vskip 5mm

\noindent {\bf Figure 9.} Variation of the proportion of binaries with
mass of the primary (Section~3.3 and equation~9). The upper ($t=0$) and
lower (after cluster disintegration) small open
circles denote our model (average of 20 simulations), and the solid circles are
measurements of G~dwarfs ($f_{\rm G}=0.53\pm0.08$, following Leinert et al.
1993), K~dwarfs ($f_{\rm K}=0.45\pm0.07$, following Leinert et al. 1993)
and M dwarfs ($f_{\rm M}=0.42\pm0.09$, Fischer \& Marcy 1992). The large open
circle is derived from the 5.2~pc stellar sample (Kroupa et al.
1993). Horizontal `errorbars' represent the primary star mass range.

\vskip 5mm

\noindent {\bf Figure 10.} The orbital parameters of triple and quadruple
systems (Section~3.4). {\bf Upper panel}: the eccentricity--period diagram for
outer
orbits. The thick dashed line and the scale of axis are as in Fig.~1.
{\bf Middle panel}: the distribution
of mass ratios. {\bf Lower panel}: The distribution of the ratios of inner to
outer
semi-major axis. Stars with three rays denote triple systems and stars with six
rays denote quadruple systems.

\vskip 5mm

\noindent {\bf Figure 11.} (Section~3.4) The $e_{\rm
outer}$--log$_{10}\left(a_{\rm
inner}/a_{\rm outer}\right)$ diagram for triple systems ({\bf upper panel}) and
quadruple systems ({\bf lower panel}). The shaded region in the upper panel
constrains the outer
orbital parameters of Proxima~Cen--$\alpha$~Cen~A/B if it is a bound triple
system. Each quadruple system has two inner orbits for one outer eccentricity.

\vskip 5mm

\noindent {\bf Figure 12.} (Section~4) {\bf Top panel}: The overall mass ratio
distribution
$q=m_2/m_1\le1$,
where $m_1$ and $m_2$ are the mass of the primary and secondary, respectively.
The initial distribution, with both component masses chosen at random from the
KTG(1.3) mass function and after
pre-main sequence eigenevolution, i.e. feeding (Section~2.2), is shown as the
dashed histogram. After the dominant mode cluster
has disintegrated we obtain the
distribution shown as the thick solid histogram.
A small number of systems appear with
$q>1$ because of our binning. These have $q=1$ and result from our feeding
hypothesis in Section~2.2. The data are tabulated
in Table~1. {\bf Lower panel}: The stellar mass function of primaries
initially after pre-main sequence eigenevolution (short dashed histogram)
and after cluster disintegration (solid
histogram) is contrasted to the stellar mass function of secondaries initially
after pre-main sequence eigenevolution
(dotted histogram) and after cluster dissolution (thick long dashed histogram).
On adding
the mass function of the secondaries and primaries at any instant we obtain
approximately the KTG(1.3) mass function.
The model distributions are averaged from 20 simulations.

\vskip 5mm

\noindent {\bf Figure 13.}
The initial distribution of specific angular momenta of our binary star
population (solid
histogram, normalised to unit area) is compared with the distribution of
specific angular momenta in molecular cloud cores (dashed histogram,
normalised to unit area) obtained from table~2 of Goodman et al. (1993). The
binary star
specific angular momentum distribution decays to the form shown as the dotted
histogram after cluster dissolution which corresponds to that for main sequence
binary stars in the Galactic field. The model distributions are averaged from
20~equivalent simulations (Section~4).

\vfill
\bye